\documentclass[superscriptaddress,prb,preprint,amsmath,amssymb]{revtex4}
\usepackage{graphicx,natbib}

\newcommand{\RM }[1]{\mathrm{#1}}

\def\kB{ k_{\RM{B}} }

\def\sx{{ s^{\RM{ex}} }}

\def\DGR{{(1-D/D_0)^{GR}}}
\def\sex{ {s^{ex}} }

\begin{document}

\title{Generalizing Rosenfeld's excess-entropy scaling to predict long-time diffusivity in dense fluids of Brownian particles: From hard to ultrasoft interactions\\}

\author{Mark J. Pond} %\email{mjp736@che.utexas.edu} 
\affiliation{Department of Chemical Engineering,
  The University of Texas at Austin, Austin, TX 78712.}

\author{Jeffrey R. Errington} %\email{jerring@buffalo.edu}
\affiliation{Department of Chemical and Biological Engineering,
  University at Buffalo, The State University of New York, Buffalo,
  New York 14260-4200, USA}

\author{Thomas M. Truskett} \email{truskett@che.utexas.edu}
\thanks{Corresponding Author} {}
\affiliation{Department of Chemical Engineering, The University of Texas
  at Austin, Austin, TX 78712.}

\begin{abstract}
Computer simulations are used to test whether a recently introduced generalization of Rosenfeld's excess-entropy scaling method for estimating transport coefficients in systems obeying molecular dynamics can be extended to predict long-time diffusivities in fluids of particles undergoing Brownian dynamics in the absence of interparticle hydrodynamic forces.  Model fluids with inverse-power-law, Gaussian-core, and Hertzian~pair interactions are considered.    Within the generalized Rosenfeld scaling method, long-time diffusivities of ultrasoft Gaussian-core and Hertzian particle fluids, which display anomalous trends with increasing density, are predicted (to within 20\%) based on knowledge of interparticle interactions, excess entropy, and scaling behavior of  simpler inverse-power-law fluids.
\end{abstract}
\maketitle

Dense suspensions of hard-sphere colloids have long served as useful experimental models  for  exploring how constraints of particle packing affect the properties of condensed phases.\cite{Pusey86,Weeks2000Three-Dimension,Lowen2000Fun-With-Hard-S} Nonetheless, many technologically relevant suspensions contain high concentrations of ``particles'' (e.g., micelles, microgel colloids, dendrimers, or star polymers) which interact with one another through considerably softer \emph{effective} potentials.\cite{Likos2001effective-interactions,Louis2000Mean-field-flui,Archer2001Binary-Gaussian}        Such systems are of fundamental interest because they exhibit novel properties that cannot be readily understood in terms of concepts derived from idealized hard-particle systems like ``excluded volume", ``crowding", ``collision rate'', etc. Examples highlighted in theoretical and computer simulation studies include reentrant and cluster crystalline phases,~\cite{Stillinger1976Phase-transitio,Watzlawek1999Phase-Diagram-o,Lang2000Fluid-and-solid,Likos1998Freezing-and-cl,Likos2001Criterion-for-d,Prestipino2005Phase-diagram-o,mladek2006-Formati,LoVersa2006-Star,Zachary2008Gaussian-core-m,Moreno2007-Cluster,Mladek2007-Clus} as well as anomalous dependencies of fluid-phase thermodynamic,~\cite{Mausbach2006Static-and-dyna,Saija2006Evaluation-of-p,Krekelberg2009Gaussian-dynamics,Krekelberg2009Generalized-Rosenfeld} structural,~\cite{Foffi2003Structural-Arre,Wensink2008Long-time-self-,Krekelberg2009Gaussian-dynamics,Pond2009-Composition-and-conc,Pamies2009Phase-diagram,Shall2010Structural-and,Jacquin2010-Anomalou,Berthier2010arXiv-Increasing} and dynamic~\cite{Foffi2003Structural-Arre,Mausbach2006Static-and-dyna,LoVersa2006-Star,Wensink2008Long-time-self-,Krekelberg2009Gaussian-dynamics,Pond2009-Composition-and-conc,Krekelberg2009Generalized-Rosenfeld,Shall2010Structural-and,Pamies2009Phase-diagram,Berthier2010arXiv-Increasing} properties on particle concentration.

To date, much of the theoretical work on these  systems has been devoted to developing and testing  statistical mechanical approaches  for predicting their structural and thermodynamic behavior.\cite{Likos2001effective-interactions,Schmidt1999-An-ab-initio-dens,Likos2006-soft,Jusufi2009-Colloq} However, some of the focus is beginning to shift toward  predicting dynamics.   For example, it has recently been shown that generalized Langevin  theories\cite{Wensink2008Long-time-self-,Medina-Noyola1988Long-time-self,Dean2004-Self} can qualitatively (but not yet quantitatively) predict the  anomalous density dependence of the long-time self-diffusivity observed in Brownian dynamics simulations of the Gaussian-core model.     Likewise, mode-coupling theory has been able to   capture some of the non-monotonic dynamic trends displayed by fluids of particles that interact via star-polymer-like,\cite{Foffi2003Structural-Arre} harmonic,\cite{Berthier2010-harmonic} Hertzian,\cite{Berthier2010arXiv-Increasing} or Gaussian-core~\cite{Shall2010Structural-and,Ikeda2010Glass-transition} pair potentials.   Kinetic theory has also been used to gain insights into the nontrivial temperature- and density-dependent trends in the long-time molecular dynamics of particles with bounded, penetrable-sphere interactions.\cite{Suh2010-molecular} 

Another promising, albeit more heuristic, approach for predicting the dynamic properties of soft-particle fluids is  a recently proposed generalization\cite{Krekelberg2009Generalized-Rosenfeld} of the excess-entropy scaling method of Rosenfeld.\cite{Rosenfeld1977Relation-betwee,Rosenfeld1999A-quasi-univers}    Here, excess entropy~$\sx$ refers to the difference between the entropy per particle of the fluid and that of an ideal gas of particles with the same  number density~$\rho$.   Excess entropy is a negative quantity, and its magnitude characterizes the extent to which static interparticle correlations -- present due  to interparticle interactions -- reduce the number of microstates accessible to the fluid.  The qualitative expectation, corroborated by data from both computer simulations  and experiments (see, e.g., refs.~\onlinecite{Rosenfeld1999A-quasi-univers,Dzugutov1996A-univeral-scal,Sharma2006Entropy-diffusi,Mittal2007Relationships-b,Abramson2007Viscosity-of-wa,Abramson2008Viscosity-of-ni,Abramson2009-Viscos,Gnan2010Pressure-en}),  is that changes to a fluid that strengthen its interparticle correlations also result in slower dynamic relaxation processes.       

In the \emph{generalized} Rosenfeld (GR) scaling approach,\cite{Krekelberg2009Generalized-Rosenfeld} information about interparticle interactions is used to recast the species' long-time tracer diffusivities\ as dimensionless (GR-reduced) combinations that are, by construction, single-valued functions of $\sx$ in the low $\rho$ limit.     For fluids of particles that interact solely via an inverse-power-law pair (IPL) repulsion, one can easily show that   GR-reduced  diffusivities remain single-valued functions of $\sx$ for all values of $\rho$ and temperature $T$. Interestingly, Krekelberg et al. have further discovered that GR-reduced tracer diffusivities obtained  from  \emph{molecular dynamics} simulations of more complex model systems --  equilibrium fluid mixtures of (additive or non-additive) hard-sphere or  Gaussian-core particles at different values of $\rho$, $T$, and composition --  approximately follow a similar excess-entropy based scaling relation.\cite{Krekelberg2009Generalized-Rosenfeld}   From a practical viewpoint, this means that knowledge of the interparticle interactions and the thermodynamic excess entropy for these model fluids allows one to predict key aspects of their long-time dynamic behavior.  

Can the GR scaling method be extended  to treat systems with different types of dynamics (e.g., with  equations of motion that incorporate dynamic effects of ``solvent'' surrounding the particles)?  In this work, we take a step toward addressing this question.  Specifically, we present simulation data that tests whether the GR scaling method can be recast in a manner useful for predicting long-time tracer diffusivities of suspended particles undergoing Brownian (i.e., overdamped Langevin) dynamics in the absence of interparticle hydrodynamic forces.  The  model fluids considered here comprise particles that interact via pair potentials $\mathcal{V}(r)$ of the IPL~[$\mathcal{V}(r)=\epsilon(\sigma/r)^{\mu}$], Gaussian-core~[$\mathcal{V}(r)=\epsilon \exp\{-(r/\sigma)^2\}]$, and Hertzian~[$\mathcal{V}(r)=\epsilon (1-r/\sigma)^{5/2}$] forms.  Here $\epsilon$ and $\sigma$ represent characteristic energy and length scales of these interactions, respectively, and $\mu$ determines the ``softness" of the IPL repulsion.  As we discuss below, a key result from our study is that GR-reduced tracer diffusivities, sampled across a wide range of $T$ and $\rho$ for these models,  approximately collapse onto a single curve when plotted versus $\sx$.  A consequence is that the tracer diffusivities of the ultrasoft Gaussian-core and Hertzian particle fluids, which display anomalous trends with increasing~$\rho$,\cite{Wensink2008Long-time-self-,Pamies2009Phase-diagram} can be estimated (to within 20\% relative error) -- with no adjustable parameters-- from knowledge of (1) the interparticle interactions, (2) the value of $\sx$ at the state point of interest, and (3) the GR scaling behavior of  simpler IPL fluids.

 For the case of Brownian dynamics considered here, it is convenient to express the GR-reduced long-time diffusivity  as\cite{Krekelberg2009Generalized-Rosenfeld}     
\begin{equation}
  \label{eq:DGR}
  \DGR \equiv (1-D/D_0) \frac{B[1- d \ln B / d \ln \beta]}{D_2},
\end{equation}
where $B$  is the second virial coefficient, $\beta \equiv 1/\kB T$, $\kB$ is Boltzmann's constant, and  $D_2=-(\partial[ D/D_0]/\partial \rho)_T|_{\rho=0}$  quantifies how pair interactions at low particle concentration modify  the tracer
diffusivity $D$ relative to the  infinite dilution value $D_0 $.  As has been shown elsewhere,\cite{Krekelberg2009Generalized-Rosenfeld} $\DGR=-\sx/\kB $ for densities low enough that $\mathcal {O}(\rho^2)$ quantities can be neglected.\

Based on eq.~\ref{eq:DGR}, it is evident that four quantities must be computed to test the GR excess-entropy scaling relation for state points across the $T$-$\rho$ plane: $B$,  $D_2$, $D/D_0 $, and $\sx$.     The first two depend solely on $\beta$ and $\mathcal{V}(r)$ and can be calculated from simple theoretical considerations.  Specifically, $B$ can be obtained by integrating the following expression numerically: $B=-(1/2) \int d {\bf r} \{\exp[-\beta \mathcal {V}(r)] -1\}$.   Likewise, $D_2$ can be computed via a one-dimensional numerical integration,   
\begin{figure}[t]
        \includegraphics[clip]{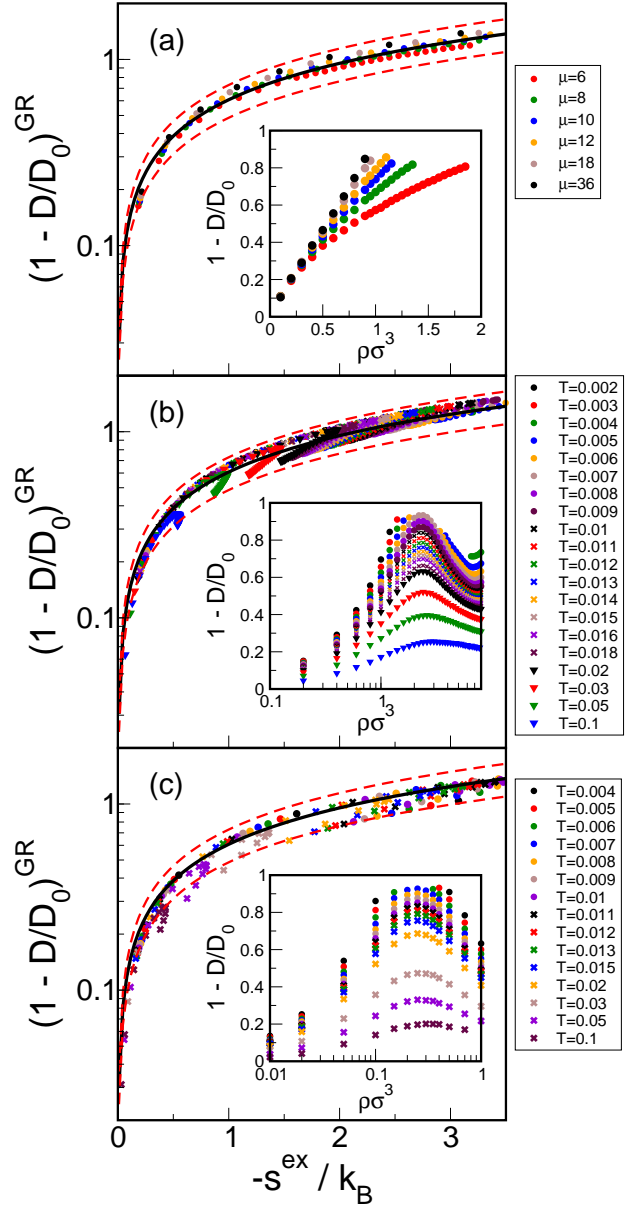}
        \caption{ GR-reduced long-time diffusivity, $\DGR$, versus negative excess entropy $-\sex / k_B$ for (a) IPL, (b) Hertzian  and (c) Gaussian-core fluids. The IPL fluids have exponents $\mu$ discussed in the text.  The solid black curve is a least-squares fit to the IPL data from Figure~\ref{fig:SSDGR}a. The dashed red lines indicate a difference of 20\% from the IPL fit. The insets show the reduced diffusivity change, $1-D/D_0$, as a function of density, $\rho \sigma^3$. }
        \label{fig:SSDGR}
\end{figure}

\begin{equation}
        D_2 =  \textbf{}\ \frac{2\pi}{3} \int_0^\infty  \frac{d[\exp(- \beta \mathcal{V}(r))]}{dr} Q(r)  r^2\ dr 
        \label{eq:Frelax2}
\end{equation} Eq.~\ref{eq:Frelax2} is derived by  the relaxation method, an approach outlined in detail elsewhere.\cite{Batchelor1976Brownian-Diff, Batchelor1983Diffusion-in-a-dil, Lekkerkerker1984On-the-calc,Dhont-book} The function $Q(r)$ is calculable from knowledge of $\mathcal{V}(r)$ via\footnote{In this work, we used the MATLAB function bvp4c to solve eq.~\ref{eq:Qofr}} 
\begin{equation}
        r^2 \frac{d^2Q}{dr^2} +r \frac{dQ}{dr}\left[ 2-r\frac{d\mathcal{(\beta V})}{dr} \right]-2Q= r^{2}\frac{d(\beta\mathcal{V})}{dr}
        \label{eq:Qofr}
\end{equation} The physical significance of $Q(r)$ is that it characterizes the extent to which application of a small external force ${\bf f}$ to a particle in the low-$\rho$ fluid would distort its pair correlation function with its neighbors, $g({\bf r}) = \exp[-\beta \mathcal{V}(r)]\{1+\beta Q(r){\bf r} \cdot {\bf f}/2r\}$, and hence modify its effective mobility.  One boundary condition on $Q(r) -$ valid for all $\mathcal{V}(r)$ considered here -- is that the force-induced distortion of the pair correlation function must decay at large distances from the central particle, $Q(\infty)=0$. The second boundary condition depends on the form of the short-range interparticle repulsion.  For interactions studied previously with a hard-sphere repulsion at a separation $r=\sigma$,\cite{Batchelor1976Brownian-Diff, Batchelor1983Diffusion-in-a-dil, Lekkerkerker1984On-the-calc,Dhont-book} the following condition holds: $(dQ/dr)|_{r=\sigma}=-1$.  We have verified that this condition also accurately describes the behavior of particles with softer IPL repulsions if it is applied at an effective hard-sphere ``Boltzmann''\cite{Ben-Amotz2004Reformulation-o} diameter $r=\sigma_{\text{B}}$ defined by $\mathcal{V}(\sigma_{\text{B}})= 10\kB T$.  To our knowledge, this work is the first to consider $D_2$ for ultrasoft potentials like the Gaussian-core or Hertzian models that are bounded at zero interparticle separation.  From eq.~\ref{eq:Qofr}, it is apparent that the condition $Q(0)=0$ applies for such models.

To compute $D/D_0,$ we use Brownian dynamics simulations, where particle translations are governed by the Langevin equation, solved in the overdamped limit. The algorithm  we employ, wherein the position ${\bf r}_i$ of each particle $i \in [1,N]$ updates at each time step according to   ${\bf r}_i(t+\Delta t)={\bf r}_i(t) - D_0\Delta t\nabla \sum_{j \neq i}^N \beta \mathcal{V}(r_{ij}(t))+\Delta {\bf r}_i$, is detailed elsewhere. \cite{Ermak1975A-Computer-Sim,Allen1987Computer-Simula} The quantity $\Delta {\bf r}_i$ denotes a random displacement due to solvent collisions; its magnitude is taken to be Gaussian-distributed with a mean of zero  and a variance of $6D_0\Delta t$.    

The Brownian dynamics simulations track $N$ particles in a periodically-replicated cubic simulation cell whose volume~$V$ sets $\rho=N/V.$  We use $N= 1000$, $2000$, and $3000$ particles for the IPL, Hertzian, and Gaussian-core fluids, respectively.    Interparticle potentials were truncated at $r=1.4 \sigma$ - $3.9 \sigma$,\footnote{The cutoff radius $r_{cut}$ was determined by $ \int_0^{r_{cut}} d {\bf r} \{\exp[-\mathcal {V}(r)/\kB T] -1\} = 0.99  \int_0^\infty d {\bf r} \{\exp[-\mathcal {V}(r)/\kB T] -1\}$} $\sigma$, and $3.71\sigma$, respectively, for these systems.   For simulations of the IPL fluids, we set $D_0 = 0.001 (\sigma^2 \epsilon / m)^{1/2}$ and the  $\Delta t = 0.025 \tau_B$, where $\tau_B \equiv m D_0/k_B T$ is a characteristic Brownian time scale. For  simulations of Gaussian-core and Hertzian particle systems, we set $D_0 = 0.0001 (\sigma^2 \epsilon / m)^{1/2}$ and $\Delta t = 0.1 \tau_B$.  Long-time tracer diffusivities are computed  from the mean-squared displacements  via the Einstein relation, $D = \langle\Delta r^2 \rangle / 6 t$ as $t\rightarrow \infty$.

We employ free-energy-based simulation techniques to evaluate $\sx$. In the first step, we obtain the $\rho$ dependence of the Helmholtz free energy at high $T$ using grand canonical transition matrix Monte Carlo simulation. \cite{Errington2003Direct-calculat} In the second step, we perform a canonical temperature expanded ensemble simulation \cite{Lyubartsev1992New-approach-to} with a transition matrix Monte Carlo algorithm \cite{Grzelak2010} to evaluate the change in Helmholtz free energy with $T$ at constant $\rho$. Collectively, these simulations provide values for excess Helmholtz free energy and excess energy, and hence $\sx$, at a state point of interest.  Interested readers can find additional details in our earlier papers. \cite{Chopra2010On-the-use,Chopra2010On-the2}

        \begin{figure}[t]
                \includegraphics[clip]{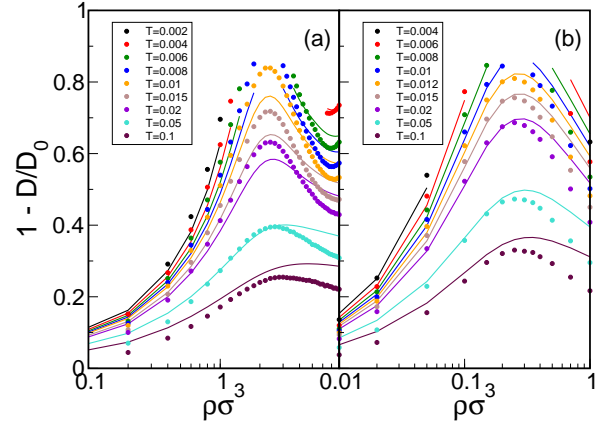}
                \caption{GR predictions (curves, discussed in text) and Brownian dynamics simulation data (symbols)  for long-time diffusivity of the (a)Hertzian and (b) Gaussian-core fluids as a function of density $\rho \sigma^3$.}
        \label{fig:Proj}
        \end{figure}

We now examine the simulated behavior of six IPL fluids with exponents of $\mu=6$, $8$, $10$, $12$, $18$, and $36$.  Specifically, Figure~\ref{fig:SSDGR}a shows how $(1-D/D_0)$ and its GR-reduced counterpart $\DGR$ vary with density $\rho \sigma^3$ and $-\sx/\kB$, respectively, for these fluids at $T=\epsilon/\kB$.  As should be expected, systems with softer interparticle repulsions (i.e., lower $\mu$) display considerably faster single-particle dynamics at a given $\rho \sigma^{3}$.  For example, the $\mu=6$ fluid must realize a density twice that of the $\mu=36$ fluid for both systems to exhibit $D/D_0=0.2$.  In contrast,  $\DGR$ for these fluids shows only minor systematic variation when plotted versus   $-\sx/\kB$, with less than 20\% difference between the $\mu=6$ and $\mu=36$ values.         % we look at the generalized Rosenfeld scaled diffusion as a function of excess entropy found in the rest of Figure \ref{fig:SSDGR}.  This comparison of a thermodynamic structural metric has a much stronger correlation with the diffusivity.  It should also be noted that the nondimensionalizeing coefficient for the scaling, $(B+T\frac{dB}{dT})/D_2$, is approximately constant across all of the exponents tested in this experiment.

To put to the GR scaling relationship for Brownian dynamics to a more stringent test, we further analyze the behaviors of model fluids with ultrasoft Hertzian and Gaussian-core interactions (see Figures~\ref{fig:SSDGR}b and \ref{fig:SSDGR}c, respectively). In particular, we explore a broad range of $\rho \sigma^3$ for these two systems along isotherms spanning from the moderately supercooled liquid ($T \le 0.004 \epsilon/\kB$) to temperatures high enough  ($T \ge 0.01 \epsilon/ \kB$) for the models to exhibit mean-field-like behavior.\cite{Pamies2009Phase-diagram,Stillinger1976Phase-transitio,Louis2000Mean-field-flui,Likos2001effective-interactions} Note that this data set includes state points  for which $D/D_0$ decreases with increasing $\rho \sigma^3$ in both systems, behavior that is anomalous when compared to the trends exhibited by simpler atomic and colloidal fluids.  The main conclusion to be drawn from Figures~\ref{fig:SSDGR}b and \ref{fig:SSDGR}c is that, despite the anomalous relationship between single-particle dynamics and $\rho$ in these systems, the GR excess entropy scaling relation looks strikingly similar to the quasi-universal form of the IPL fluids.  In fact, the  Hertzian and Gaussian-core data deviate by less than 20\% from the least-squares fit to the $\DGR$ versus $\sx$ data from Figure~\ref{fig:SSDGR}a.

One consequence of this quasi-universal scaling behavior is that one can estimate Brownian dynamics $D/D_0$ data, \emph{even for nontrivial systems with ultrasoft potentials}, simply from knowledge of their interparticle interactions, the excess entropy, and, e.g., a least-squares fit of the IPL data of Figure~\ref{fig:SSDGR}a.  Figure~\ref{fig:Proj} illustrates the quality of the predictions made by this approach for the Hertzian and Gaussian-core model fluids.    As can be seen -- with no adjustable parameters -- the estimated values of diffusivity are accurate for low-to-moderate $\rho$, and remain qualitatively reliable for higher $\rho$ where the anomalous trends emerge.\   Collectively, these results illustrate a strong apparent link between static interparticle correlations and long-time dynamics for model fluids with either hard or ultrasoft interactions. 

Finally, whether the approach outlined here can provide an accurate estimate to the ``thermodynamic'' component of diffusivity for systems which also have strong hydrodynamic interparticle contributions to dynamics is an  open question which deserves attention in future studies.  Furthermore, it will be interesting to test whether the GR scaling for diffusivity in Brownian dynamics also holds for systems with more complex interparticle interactions (e.g., with more than one characteristic length scale).

We thank Prof. David S. Dean for suggesting the relaxation method for computing~$D_2$. \ T.M.T. acknowledges support of the
Welch Foundation (F-1696) and the David and Lucile Packard Foundation.
J. R. E. acknowledges financial support of the National Science Foundation, Grant No. CBET-0828979.  M.J.P. acknowledges the support of the Thrust 2000 - Harry P.
Whitworth Endowed Graduate Fellowship in Engineering. The Texas
Advanced Computing Center (TACC), the University at Buffalo Center for Computational Research, and the Rensselaer Polytechnic Institute Computational Center for Nanotechnology Innovations provided computational resources for
this study.

%\bibliography{mjp_firstbib}

\end{document}